\documentstyle[aps,twocolumn]{revtex}

\def\vphi{\varphi({\bf r}, t)}

\def\be{\begin{equation}}
\def\ee{\end{equation}}
\def\beqa{\begin{eqnarray}}
\def\eeqa{\end{eqnarray}}

\def\Psian{\tilde{\Psi}({\bf r},t)}

\def\rhoo{\rho({\bf r}, t)}

\begin{document}
\title{Josephson effect between trapped Bose-Einstein condensates}

\author{
Ivar Zapata,${}^1$ Fernando Sols,${}^1$ and Anthony J.~Leggett${}^2$
}
\address{
${}^1$Departamento de F\'{\i}sica Te\'orica
de la Materia Condensada,
Universidad Aut\'onoma de Madrid, E-28049 Madrid, Spain \\
${}^2$Department of Physics, University of Illinois at Urbana-Champaign,
1110 West Green Street, Urbana, Illinois 61801
}
\maketitle

\begin{abstract}
We study the Josephson effect between atomic
Bose-Einstein condensates. By drawing on an electrostatic
analogy, we derive a semiclassical functional expression for the
three-dimensional Josephson coupling energy in terms of the condensate
density.
Estimates of the capacitive energy and of the Josephson plasma
frequency are also given. The effect of dissipation due to the
incoherent exchange of normal atoms is analysed.
We conclude that coherent Josephson dynamics may already be observable
in current experimental systems.
\end{abstract}

Recently  \cite{JILA95,TEXAS95,MIT96}, it has become possible to cool a
macroscopic number ($\sim 10^4-10^6$) of
magnetically confined, spin-polarized atoms down
to temperatures of order 100 nK while
maintaining densities sufficiently high
($10^{11}-10^{15}{\rm cm^{-3}}$) to permit the onset of Bose-Einstein 
condensation (BEC).
From the ensuing theoretical work, it has been
concluded that
these Bose condensed atomic gases behave very differently from the ideal 
noninteracting
gases, which yields the prospect of potentially displaying a rich
phenomenology that might
include vortex states and the
Josephson effect \cite{BAYM96,STRINGARI96,OXFORD96}.

In this article we present a theoretical study of the Josephson
dynamics between two atomic baths that have undergone BEC. The
Josephson effect results from a collective mode of two weakly
connected systems between which a macroscopic fraction of particles
can tunnel with identical probability amplitude. Going beyond previously
proposed
one-dimensional \cite{DALFOVO96} and dissipation free
\cite{DALFOVO96,TRIESTE97}
models, we present here a three-dimensional study of
the Josephson effect between
Bose condensates and
estimate the effect of damping. We calculate the Josephson coupling
energy, the capacitive energy (which accounts for quantum fluctuations
of the phase), and the frequency of the Josephson plasma oscillation.
Our
main conclusion is that still lower temperatures than those achieved
up to date are needed for a clear realization of the Josephson effect.

The collective dynamics of a Bose condensate at zero temperature is
described by its macroscopic wave function $\Psi ({\bf
r}, t)$. If this is factorized as $\Psi=\sqrt{\rho}\exp(i\varphi)$,
the standard energy functional can be written as
\be
\label{PhNumH}
H = \int d{\bf r} \left[ \frac{\hbar^2}{2m}\left(
|\nabla
\sqrt{\rho}|^2+
\rho|\nabla\varphi|^2
\right)
+
V_{\rm ext}\rho +g\rho^2
\right]
\ee
($g=2\pi\hbar^2a/m$), and the corresponding Hamilton equations
lead to the Gross-Pitaveskii equations
\cite{STRINGARI96,GP}. Neglecting depletion \cite{stri97}, the
normalization can be taken as $\int d{\bf r} \rho=N$, being $N$ the
total
number of atoms.

At sufficiently low temperatures the phase within one well can be
regarded as
uniform. This can be easily seen if one estimates the energy of a
one-radian
fluctuation of the phase across the condensate near equilibrium
in a single spherical harmonic well
[see Eq. (\ref{PhNumH})],
\be
\label{EFl}
\frac{\hbar^2}{2m}\int d{\bf r}\rho_{\rm eq}|\nabla \delta
\varphi|^2\simeq \frac{\hbar^2}{2m}\frac{N}{4R^2}
\simeq 0.7 N^{3/5}  \frac{ \hbar
\omega_0}{2},
\ee
where $R=a_0(15aN/a_0)^{1/5}$ is the cloud radius
estimated within the
Thomas-Fermi approximation \cite{BAYM96}, $\omega_0$ is the harmonic
oscillator frequency within the well, and $a_0=(\hbar/m\omega_0)^{1/2}$
is the oscillator length. For the last approximate equality we have
used typical parameters $a=5$ nm and $a_0=10^{-4}$ cm. The
characteristic temperature of such an oscillation can be as big as
10 $\mu$K for the MIT (1996) experiment \cite{MIT96}. The estimate
(\ref{EFl})
suggests that, in the interestitial region between two wells, where
$\rho({\bf r})$ decreases appreciably, spatial phase variations are
less costly and thus easier to activate at low temperatures. Here lies
the essence of the low energy Josephson dynamics which we study below.

Let us consider two weakly connected condensates $1$ and $2$.
We assume that the condensates are confined within spherical
harmonic wells of the same frequency $\omega_0$.
First we wish to analyse the semiclassical dynamics. It is, of course,
well known that the relative number $\delta N$ and the relative phase
$\chi$
of the condensate in the two wells may be treated as canonically
conjugate 
variables. Nevertheless, by introducing a sufficiently coarse-grained
average 
$n$ of the number in (say) well $1$ we may treat $n$ and $\chi$ as 
simultaneously well-defined, and write for the wave function the ansatz
\be
\label{ansatz}
\Psian \propto \Psi_1({\bf r}; n(t))+e^{i\chi(t)}\Psi_2({\bf
r};N-n(t)),
\ee
where $\Psi_i({\bf r}; n)$ is the (real) equilibrium wave function for
the isolated well $i$ containing $n$ bosons. It is straightforward to
show that, to lowest order in the overlap integrals $\int \Psi_1
\Psi_2$, the energy functional for $\tilde{\Psi}$ takes the form
\be
\label{RedEn}
H(\delta N, \chi)
\simeq E_B
(\delta N) + E_J(\delta N)\left(1 - \cos \chi \right),
\ee
where $E_B(\delta N)$ is the bulk energy of the two isolated wells
with $\delta N \equiv N/2-n$ transferred atoms, and $E_J(\delta N)$ is
the Josephson coupling energy. $E_B(\delta N)$ may be expanded as
\be
\label{EBulk}
E_B(\delta N)\simeq E_{B}(0)+\mu'(\frac{N}{2})\delta N^2 +
\frac{1}{12} \mu'''(\frac{N}{2})\delta N^4.
\ee
Within the Thomas-Fermi approximation, $\mu\sim N^{2/5}$ \cite{BAYM96},
so that the
ratio between the third and second terms in the expansion is $0.32 \,
\delta N^2/N^2$, which means that the last term can be neglected in a
wide range of situations. To avoid complications stemming from
possible resonances between Josephson oscillations (see below) and
intrawell excitations, we require $\mu(N/2+\delta N)-\mu(N/2-\delta
N)\ll \hbar \omega_0$, where we use the result that the first normal
mode of a spherical well lies approximately at $\hbar \omega_0$ above
the ground state \cite{STRINGARI96,OXFORD96}. This condition is
realized when $\delta N/N\ll 4.6 N^{-2/5}$ for typical parameters. This
may seem an important restriction on $\delta N$. However,
the fraction of transferred atoms can be as big as $10$\% for the
JILA experiment \cite{JILA95}, but it has to be $<2$\% for the MIT
experiment \cite{MIT96}.

The expression for the coupling energy $E_J(\delta N)$ in terms of
$\Psi_1$ and
$\Psi_2$ which can be derived from Eqs. (\ref{ansatz}-\ref{RedEn}) is
rather 
complicated and
difficult to handle. A much simpler expression can be obtained if
one notes that the coupling energy must come entirely from the
$\nabla\varphi$
term in
Eq.(\ref{PhNumH}), which in turn can be approximated as
\be
\label{PhH}
\int_{\rm ext} d{\bf r} \frac{\hbar^2}{2m}\rhoo|\nabla\vphi|^2  \equiv
E[\varphi],
\ee
where the integration extends over the the region exterior to the
condensates (the results are quite independent on the precise location
of the condensate borders). As argued before, the phase within the
condensates can be assumed to be practically uniform. In the limit of
very small phase difference, $\chi \ll 1$, the phase term in Eq.
(\ref{RedEn}) can be approximated as $E_J\chi^2/2$. The only way for
Eq. (\ref{PhH}) to have such a dependence on the total phase
difference $\chi$ and not on the details of $\nabla\varphi$ is that
the condition $\delta E[\varphi]/\delta
\vphi=0$ is satisfied. One may note that, except for
trivial factors, this is the electrostatic equation for the electric
displacement vector ${\bf D} \equiv
\rho \nabla \varphi$ in a medium with a nonuniform dielectric constant
$\rhoo$. Boundary conditions for $\vphi$ are given by its value at the
borders of each condensate, which act as conductors in this analogy.
We have a system of two conductors held at a potential difference
$\chi$ and a dielectric medium surrounding them. Then $E[\varphi]$ is
essentially the energy of this capacitor. Potential theory tells us
that $E[\varphi] \simeq (\hbar^2/2m) C[\rho] \chi^2$, which implies
\be
\label{EJ}
E_J= \frac{\hbar^2}{m} C[\rho],
\ee
where $C[\rho]$ is the
mutual capacitance of the two conductors in the presence of such a
dielectric \cite{note3}.

We take the origin of coordinates in the middle point of the
double well configuration, taking the $z$ direction along the line
which joins the two minima. Standard variational arguments
can be invoked to prove that,
for parallel plate boundary conditions,
\be
\label{EJCond2}
\int\frac{dxdy}{\int_1^2 dz \rho(x,y,z)^{-1}} \le C[\rho]
\le \left[ \int_1^2\frac{dz}{\int dxdy
\rho(x,y,z)}\right]^{-1}. 
\ee
The lower bound is obtained by removing the positive term
$(\partial\varphi/\partial x)^2+(\partial\varphi/\partial y)^2$ from the
energy
functional (\ref{PhH}), while the upper bound is derived by taking
$\varphi$ independent of $x,y$. If $V_{\rm ext}$ depends weakly on $x,y$
in
the region that controls the capacitance, we can write
$\rho(x,y,z)\simeq
\rho(0,0,z)$ and then the two bounds become approximately equal to
\be
\label{EJCondensator}
C[\rho] \simeq A \left[ \int_1^2\frac{dz}{\rho(0,0,z)}\right]^{-1},
\ee
where $A$ is an effective area.

In order to proceed further, we need an estimate of $\rho=|\Psi|^2$
in the region of
interest. For two 
identical wells in
equilibrium, the
ground state wave function is symmetric in $z$. Neglecting the
dependence
on $x, y$,  and within the WKB approximation \cite{DALFOVO96}, we can
write
\be
\label{psiwbk}
\Psi(z)=\frac{B}{\sqrt{p(z)}}\cosh\left[\frac{1}{\hbar}\int_0^zdz'
p(z')\right]
\ee
where $p(z)=[2m(V_{\rm ext}(0,0,z)-\mu)]^{1/2}$ and $B$ is a
constant to be determined later. Introducing (\ref{psiwbk}) into
(\ref{EJCondensator}) we obtain
\be
E_J \simeq \frac{\hbar A |B|^2}{m}
\left[2\tanh\left(\frac{S_0}{2}\right)\right]^{-1},
\ee
where $S_0\equiv\int_1^2p(z)dz/\hbar$.
For
standard (quartic) barriers, $S_0\simeq 2 \pi \alpha
(V_0-\mu)/\hbar\omega_0$, where $V_0$ is the barrier height along the
line $x=y=0$,
and $\alpha$ is of order unity ($\alpha=8/3\pi$ for $\mu\ll V_0$ and
$\alpha=1/\sqrt{2}$ for $V_0-\mu\ll V_0$).

The coefficient $B$ must be calculated by properly matching
(\ref{psiwbk}) with good approximate solutions near the border of each
well. If $S_0\agt 1$, then Eq.
(\ref{psiwbk}) can be matched with the solution given in Ref.
\cite{DALFOVO96} for the region $R \gg z-R \gg d$,
$d=a_0\left(a_0/2R\right)^{1/3}$ being the distance from the classical
radius $R$ where the Thomas-Fermi approximation begins to fail. The
result is $B\simeq u e^{-S_0/2} 
(\hbar/8\pi a d^3)^{1/2}$, with $u\simeq 0.397$ \cite{DALFOVO96}.
To estimate the effective area $A$ we note \cite{DALFOVO96} that the
form of the order parameter has a universal dependence on the position
near the boundary of the Thomas-Fermi zone, namely $\Psi(r)\simeq
\phi\left[(r-R)/d\right]/d\sqrt{8\pi a}$, where $\phi$ is a universal
function (in that it does not depend on the confining potential) which
varies appreciably within a scale of unity. Therefore $\rho(x,y,z)$
varies transversally on a scale such that $(\sqrt{x^2+y^2+R^2}-R)/d
\sim 1$. These considerations yield an estimate of
$A=2^{2/3}v(a_0/R)^{4/3}\pi R^2$, where $v\sim 1$. The
result is
\be
\label{EJ2}
E_J\simeq \frac{5.95 u^2 v e^{-S_0}}{\tanh (S_0/2)}
\left(\frac{N}{2}\right)^{1/3}\left(\frac{15a}{a_0}\right)^{-2/3}
\frac{\hbar
\omega_0}{2}.
\ee
We note that, because the prefactor in (\ref{EJ2}) has the 
same $N^{1/3}$ explicit
dependence as the critical temperature $T_c$ for a
system of free confined bosons (which is approximately equal
to the critical temperature of interacting bosons \cite{strin96}), one
could 
expect a simple relation
between the two magnitudes. Knowing that $k_B
T_c \simeq N^{1/3}\hbar\omega_0$ \cite{strin96}, then (\ref{EJ2})
can be rewritten as (for typical cases)
\be
\label{EJ3}
E_J \sim k_B T_c e^{-S_0}.
\ee

Also within the Thomas-Fermi approximation, a simple expression can be
obtained for $E_{B}(\delta N)$ in Eq.(\ref{EBulk}), which, up to
quadratic order in $\delta N$, can be written $E_{B}(\delta N)-E_{B}(0)=
E_C\delta N^2/2$, where $E_C=2\mu'(N/2)$ is the capacitive energy. From
the 
result of Ref.
\cite{BAYM96} for $\mu(N/2)$, we obtain
\be
\label{EC}
E_C\simeq \frac{4}{5}\left(\frac{N}{2}\right)^{-3/5}
\left(\frac{15a}{a_0}\right)^{2/5}\frac{\hbar \omega_0}{2}.
\ee
Collecting terms, the Hamiltonian (\ref{RedEn}) can be written
\be
\label{JosPen}
H\simeq \frac{E_C}{2}\delta{N}^2 + E_J(1-\cos\chi),
\ee
giving the well-known equivalence to a pendulum \cite{ANDE65}.
The frequency of small oscillations (the
Josephson plasmon), $\omega_{JP}=\sqrt{E_JE_C}/\hbar$, turns out to be
\be
\label{JosPlas}
\omega_{JP} \simeq 1.54 u\sqrt{v}
\left(\frac{2a_0}{15aN}\right)^{2/15}
\frac{e^{-S_0/2}}{\sqrt{\tanh (S_0/2)}} \,\, \omega_0,
\ee
which is typically a fraction (not necessarily very small) of the
confining potential. As usually $\omega_0/2\pi \simeq 10-100$ Hz, we
conclude 
that, $\omega_{JP}\alt 10$ Hz.

The ratio $E_J/E_C$ is  a good measure of the classical character of the 
relative phase $\chi$.
From (\ref{EJ2}) and (\ref{EC}), we find
\be
\frac{E_J}{E_C}\simeq 0.25 u^2 v \left(\frac{2a_0}{15aN}\right)^{1/15}
\frac{e^{-S_0}}{\tanh(S_0/2)}
\frac{Na_0}{a}.
\ee
By varying $S_0$, the system can be driven from the classical regime
($E_J\gg E_C$) to the strong quantum limit ($E_J\ll E_C$). However, in
the latter case, quantum fluctuations are only important if  we operate
at ultralow temperatures $k_BT \alt E_C$.

The Hamiltonian (\ref{JosPen}) describes the dynamics of a
conservative system. In real life, however, we should expect
a certain amount of 
damping. The
most obvious source of such damping
is the incoherent exchange of normal atoms, and a quantitative
discussion requires a generalization of our results to
nonzero temperature. Because the spatial scale of the normal component
is quite different, in a harmonic trap, from that of the condensate,
this generalization is much less trivial than for the case of a
junction linking two homogeneous superconductors or superfluids, and we
shall not attempt a quantitative discussion here.
However, we shall give two qualitative arguments, based on two very
different 
physical
assumptions (corresponding essentially to the high and low barrier
limits), to 
the effect that the
damping associated
with the normal component will overdamp the Josephson behavior and
hence make it in practice unobservable down to temperatures of the
order of $\hbar\omega_0/k_B$.  In both cases, we find that normal
atoms give an 
Ohmic
contribution to the current,
\be
\label{Ohmic}
I_n=- G \delta \mu,
\ee
where $\delta \mu$ is the chemical potential difference (for simplicity,
we
assume equilibrium within each well).
As a result, the first
Josephson equation is modified to read
\be
\label{disip2}
\frac{d}{dt} \delta N = \frac{E_J}{\hbar} \sin \chi  +I_n.
\ee

In the high barrier limit ($V_0 \gg k_B T$), the basic
order-of-magnitude assumption is that any power-law factors occurring
are negligible compared to the relevant WKB or Arrhenius-Kramers
exponentials. At temperatures $T \alt T_c$, the normal component in
each well will be appreciable and will be distributed over an energy
range $\sim k_BT \gg \hbar \omega_0$, since \cite{stri97}
$k_BT_c/\hbar\omega_0 \sim N^{1/3}\gg 1$. We will assume that $k_BT$
is, nevertheless, small compared to $V_1 \equiv V_0-\mu_0$, with
$\mu_0
\equiv \mu(N/2)$. The situation we have is $V_0 \gg k_B T \sim \mu_0
\gg
\hbar \omega_0$. Under these conditions the typical spacing of the
one-particle energy levels at energies $\sim k_BT$ is small compared
to $\hbar\omega_0$, and we can ignore
``level-crossing'' effects \cite{LevCros} and treat the tunneling
of uncondensed 
particles as incoherent.
At high $T$,
the total rate of crossing is then given by a standard Arrhenius-Kramers 
formula, $P_n= (\kappa \omega_0/2\pi)\exp(-V_1/k_BT)$,
where usually \cite{HANGGI90} $\kappa \sim 1$. We have, approximately,
$G \simeq  P_nN_n/k_BT$
where $N_n$ is the number of normal particles. The
degree of damping can be estimated by comparing
 (\ref{Ohmic}) with the Josephson supercurrent in a small
oscillation of amplitude $\delta \mu$, namely,
\be
\label{disip4}
I_s \sim \frac{E_J \delta \mu}{\hbar^2 \omega_{JP}}.
\ee
From (\ref{JosPlas}), we estimate
$\omega_{JP} \sim N_0^{-2/15} e^{-S_0/2} \omega_0$
($N_0 \sim N$ is the number of
condensed particles),
valid for typical parameters at $T=0$. Finite $T$ corrections should
not change the qualitative conclusions. We find
$I_s/I_n \sim 2\pi N_0^{7/15} (\hbar\omega_0/k_B T)^2
 \exp (V_1/k_BT - S_0/2)$,
since $N_n/N \sim (T/T_c)^3$. 
We conclude that, in the high-barrier limit, $I_s/I_n \ll 1$,
hence the motion of
the equivalent pendulum is overdamped, at least
down to temperatures of the order of $\hbar\omega_0/k_B$. For
temperatures below this our estimate fails for several reasons, not
least because the density of normal component falls exponentially
rather than as $T^3$.

The estimate presented above is, however, quite irrelevant for today's
experiments, because the high barrier condition requires $S_0/\pi \gg 2
\mu_0/\hbar \omega_0 \sim 6 - 100$ (we take $\alpha\simeq 1$)
for typical situations. On
the other hand, the prefactor in $E_J/\hbar$ can be estimated
as $10^4-10^5$ Hz, and thus the critical current itself would be so
small as to
be completely unobservable. A more relevant
regime is that in which the chemical potential lies near the top of
the barrier, $V_1\sim\hbar\omega_0/2$, so that $S_0 \sim 1-5$. In this
regime, we can still expect the WKB formula for $E_J$ derived above to
yield a reasonable approximation. However, if
$k_BT\gg\hbar\omega_0/2\sim V_1$, the thermal cloud lies mostly above
the barrier and a radically different approach is needed to study its
transport properties. For simplicity, we introduce the drastic
approximation that particles impinging on the barrier with energy $E$
are transmitted with probability one if $E>V_0$, and zero if $E<V_0$.
Then, the flow of normal atoms due to a fluctuation in $\delta \mu$ is
only limited by the ``contact resistance", a concept taken from
ballistic transport in nanostructures \cite{imry86}. Adapting standard
arguments to the case of bosons, we may write  $G\equiv N_{\rm ch}/h$,
where $N_{\rm ch}$ is an effective number of
available transmissive channels.  Within a continuum approximation,
and assuming that only transverse channels with a minimum energy
between $\mu_0$ and $\mu_0+k_BT$ are populated, we find (taking $\delta
\mu
\ll k_BT$) $N_{\rm
ch}=mA_nk_BT/2\pi\hbar^2$, where $A_n$ is a mean transverse
contact area seen by normal particles \cite{note4}.
Approximating \cite{stri97} $A_n \sim 2\pi k_BT/m\omega_0^2$, we
find $N_{\rm ch} \sim (k_BT/\hbar\omega_0)^2$, and hence 
\be
\label{ratio}
I_s/I_n
\sim 2 \pi N_0^{7/15} (\hbar \omega_0/k_BT)^2 e^{-S_0/2} ,
\ee
which, interestingly, is formally equivalent to the high barrier result
 (since there $S_0/2\sim \pi\alpha V_1/\hbar\omega_0 \gg V_1/k_BT$).
Taking $k_BT=10\hbar \omega_0$, we find $I_s/I_n \sim 0.38$ for
$N \sim 10^4$, and 3.3 for $N \sim 10^6$, if $S_0 \sim 5$. The
equivalent
numbers for $S_0 \sim 1$ are 2.8 and 24. We tentatively
conclude that effects of coherent Josephson dynamics can be observed
in today's atomic Bose condensates if the barrier is low. Underdamped
dynamics can be further favored by decreasing $T$ and increasing $N$.
It is important to note, however, that
some aspects of
the Josephson behavior can be observed in the overdamped regime,
provided that thermal fluctuations are unimportant ($k_BT/E_J \alt 1$),
\cite{ambe69} a not very stringent condition in the low
barrier limit [see Eq. (\ref{EJ3})].

In summary, a systematic study of the Josephson effect between two
weakly connected atomic Bose-Einstein baths has been presented. We
have derived a novel, three-dimensional functional expression for
the Josephson coupling energy in terms of the condensate density. The
capacitive energy and the frequency of the Josephson plasma
oscillation have also been calculated within the WKB approximation.
The effect of damping due to the incoherent exchange of normal atoms
has been estimated in the limits of high and low potential
barrier. Within the low barrier regime, we find
[see Eq. (\ref{ratio})] that weakly damped Josephson dynamics may
already be observed in current experimental setups, and that coherence
between atomic Bose condensates can be further enhanced by lowering the
temperature and the potential barrier, as well as by increasing the
number
of condensate particles.

This work has been supported by DGICyT (PB93-1248) and by NSF
(DMR 96-14133).  One of us (A.J.L.) wishes to thank the BBV
Foundation for an invited professorship at the Universidad Aut\'onoma
de Madrid, and Dr. F. Sols and his colleagues for their kind
hospitality.


\begin{thebibliography}{1}

\bibitem{JILA95} M. H. Anderson, J. R. Ensher, M. R. Matthews,
C. E. Wieman and
E. A. Cornell, Science {\bf 269}, 198 (1995).

\bibitem{TEXAS95} C. C. Bradley, C. A. Sackett and J. J. Tollett and R.
G. 
Hulet,
Phys. Rev. Lett. {\bf 75}, 1786 (1995).

\bibitem{MIT96} M. O. Mewes, M. R. Andrews, N. J. van Druten, D. M.
Kurn, D.
S. Durfee, C. G. Townsend and W. Ketterle, Phys. Rev. Lett. {\bf 77},
416
(1996); {\bf 77},988 (1996).

\bibitem{BAYM96} G. Baym and C. J. Pethick, Phys. Rev. Lett. {\bf 76}, 6
(1996).

\bibitem{STRINGARI96} S. Stringari, Phys. Rev. Lett. {\bf 77}, 2360
(1996).

\bibitem{OXFORD96} M. Edwards, P. A. Ruprecht, K. Burnett, R. J. Dodd
and
C. W. Clark, Phys. Rev. Lett. {\bf 77}, 1671 (1996).

\bibitem{DALFOVO96} F. Dalfovo, L. Pitaevskii and S. Stringari, Phys.
Rev. A
{\bf 54}, 4213 (1996).

\bibitem{TRIESTE97} A. Smerzi {\it et al.}, cond-mat/9706221; S.
Raghavan
{\it et al.}, cond-mat/9706220.

\bibitem{GP} L. P. Pitaevskii, Zh. Eksp. Teor. Fiz. {\bf 40}, 646 (1961)
[Sov.
Phys. JETP {\bf 13},451 (1961)]; E. P. Gross, Nuovo Cimento {\bf 20},
454
(1961); J. Math. Phys. {\bf 4}, 195 (1963).

\bibitem{stri97} S. Giorgini, L. P. Pitaevskii, and S. Stringari,
cond-mat/9704014.

\bibitem{note3} A more direct argument goes as follows. Rewrite Eq.
(\ref{PhH}) as
$E[\varphi]=(\hbar^2/2m)\int_{\rm ext}d{\bf r} \rho (\nabla
e^{i\varphi}) \cdot (\nabla e^{-i\varphi})$, and let $Z({\bf r})$ be a
solution of $\nabla \cdot(\rho \nabla Z)=0$ with boundary conditions
$Z(j)=e^{-i\varphi_j}$, ($j=1,2$). Identifying $Z({\bf
r})=e^{-i\varphi({\bf r})}$ and integrating by parts, we obtain
$E[\varphi]=(\hbar^2/2m)C[\rho](1-\cos \chi)$. A drawback of this
reasoning is that $|Z|=1$ is not guaranteed everywhere.

\bibitem{strin96} S. Stringari, Phys. Rev. Lett. {\bf 76}, 1405 (1996).

\bibitem{ANDE65} P. W. Andersson, in {\it Lectures in the Many Body
Problem},
E. Caianello, ed. (Academic, New York, 1964).


\bibitem{LevCros} R. Rouse, S. Han and J. E. Lukens, Phys. Rev. Lett
{\bf 75},
1614 (1995).

\bibitem{HANGGI90} P. H\"anggi, P. Talkner, and M. Borkovec,
Rev. Mod. Phys. {\bf 62}, 251 (1990).

\bibitem{imry86} Y. Imry, in {\it Directions in Condensed Matter},
G. Grinstein and E. Mazenko,
eds. (World Scientific, Singapore, 1986).

\bibitem{note4}  A more accurate calculation
which exactly accounts for the thermal population of all channels (but
yet assumes an energy-independent mean value of $A_n$) would
replace $k_BT$ by 
$[V_1(e^{V_1/k_BT}-1)^{-1}-k_BT\ln(1-e^{-V_1/k_BT})]
\simeq
k_BT\ln(k_BT/V_1)$.

\bibitem{ambe69} V. Ambegaokar and B. I. Halperin, Phys. Rev. Lett.
{\bf 22}, 1364 (1969).


\end{thebibliography}
\end{document}